%% file: ms.tex
\documentclass[12pt,preprint]{aastex}
\begin{document}
\newcommand{\msun}{\mbox{M$_{\odot}$}}
\newcommand{\rsun}{\mbox{R$_{\odot}$}}
\title{WR~20a is an Eclipsing Binary: Accurate Determination of Parameters 
for an Extremely Massive Wolf-Rayet System\altaffilmark{1}}

\author{A. Z. Bonanos, K. Z. Stanek}
\affil{Harvard-Smithsonian Center for Astrophysics, 60 Garden St.,
Cambridge, MA~02138}
\affil{\tt e-mail: abonanos@cfa.harvard.edu, kstanek@cfa.harvard.edu}

\author{A.~Udalski, L.~Wyrzykowski\altaffilmark{2}, K.~\.Zebru\'n,
M.~Kubiak, M.~K.~Szyma\'nski, O.~Szewczyk,
G.~Pietrzy\'nski\altaffilmark{3}, I.~Soszy\'nski}

\affil{Warsaw University Observatory, Al. Ujazdowskie 4, 00-478
Warszawa, Poland} 
\affil{\tt e-mail: \{udalski, wyrzykow, zebrun,  mk, msz,
szewczyk, pietrzyn, soszynsk\}@astrouw.edu.pl}

\altaffiltext{1}{Based on observations obtained with the  1.3 m Warsaw 
telescope at Las Campanas Observatory, which is operated by the
Carnegie Institute of Washington.}
\altaffiltext{2}{School of Physics and Astronomy and Wise Observatory,
Tel-Aviv University, Tel~Aviv~69978, Israel}
\altaffiltext{3}{Universidad de Concepci{\'o}n, Departamento de Fisica,
Casilla 160-C, Concepci{\'o}n, Chile}

\begin{abstract}

We present a high-precision $I$-band light curve for the Wolf-Rayet
binary WR~20a, obtained as a sub-project of the Optical Gravitational
Lensing Experiment. Rauw et al. have recently presented spectroscopy
for this system, strongly suggesting extremely large minimum masses of
$70.7 \pm 4.0\; \msun$ and $68.8 \pm 3.8\; \msun$ for the component
stars of the system, with the exact values depending strongly on the
period of the system. We detect deep eclipses of about $0.4\;$mag in
the light curve of WR~20a, confirming and refining the suspected
period of $P=3.686$ days and deriving an inclination angle of
$i=74.5\pm 2.0\deg$. Using these photometric data and the radial
velocity data of Rauw et al., we derive the masses for the two
components of WR~20a to be $83.0 \pm 5.0\; \msun$ and $82.0 \pm 5.0\;
\msun$. Therefore, WR~20a is confirmed to consist of two extremely
massive stars and to be the most massive binary known with an accurate
mass determination.

\end{abstract}
\keywords{binaries: eclipsing -- binaries: spectroscopic -- stars:
Wolf-Rayet -- stars: fundamental parameters -- stars: individual:
WR~20a}

\section{Introduction}

Measuring accurate masses for the most massive stars in our Galaxy and
beyond is important for constraining star formation and stellar
evolution theories, which have indirect implications for many objects
that are not well understood, such as supernovae, gamma-ray bursts and
Population III stars. The most massive candidates in the Milky Way are
LBV 1806-20 \citep{Eikenberry04} and the Pistol Star \citep{Figer98},
which have inferred masses up to $\sim 200\;$\msun. HDE~269810 in the
LMC is another massive candidate $\sim 150\;\rm \msun$
\citep{Walborn04}. Finally, $\eta$ Carinae, one of the most studied
massive stars has an estimated mass of $\sim 120\;$\msun, yet it
remains unknown whether it is a rapid rotator or a close binary system
\citep[see review by][]{Davidson97}. However, all these masses are
only indirect estimates and thus have large uncertainties associated
with them. The only direct way of measuring accurate masses of distant
stars is through double lined spectroscopic binary systems with
eclipses present in their light curves. The limited number of existing
mass measurements for massive stars ($>20\; \msun$) can be explained
by their intrinsic rarity and the observationally demanding process of
discovering them in eclipsing binary systems.

Until recently, the most massive stars ever weighed in binaries were:
R136-38 (O3V+O6V) in the LMC with a primary mass of $56.9\pm 0.6
\;\msun$ \citep{Massey02}, WR 22 (WN7+abs$\;+\;$O), with a minimum
primary mass of $55.3\pm 7.3\; \msun$ \citep{Rauw96,Schweickhardt99}
and Plaskett's star with a minimum primary mass of $51\; \msun$
\citep{Bagnuolo92}. However, WR~20a, a Wolf-Rayet (WR) binary in the
compact cluster Westerlund~2, seems to be the new record
holder. \citet{Rauw04} obtained spectroscopy for WR~20a and measured
extremely large minimum masses of $70.7 \pm 4.0$ and $68.8 \pm
3.8\;\msun$ for the components. The final masses strongly depend on
both the period and inclination of the binary, which can only be
measured from the light curve. They derived a period of $3.675\;$days
from the absolute values of the radial velocity differences and
assumed a circular orbit to derive the masses of the components. In
their discussion of WR~20a, \citet{Rauw04} stressed the importance and
necessity of photometric monitoring of this binary.

\citet{Shara91} were the first to make the discovery that WR~20a was a
WR star, by obtaining spectroscopy of the system. \citet{Moffat91}
obtained $UBV$ photometry, however, as \citet{Rauw04} suggest, it is
likely affected by photometric eclipses. WR~20a was classified as a
candidate binary along with 30 other WR stars in the VIIth Catalogue
of Galactic Wolf-Rayet Stars \citep{vanderHucht01}, because of its
comparatively weak emission lines, which were thought to be diluted by
the continuum light of an OB-type companion. \citet{Rauw04}
spectroscopically surveyed these candidate WR binaries and WR~20a
clearly emerged as an interesting object, which they followed up. Due
to its favorable location in the sky, this object was included in the
target list covered by the Optical Gravitational Lensing Experiment
(OGLE) photometric survey. We have quickly obtained a precise light
curve of WR~20a and thus confirm and refine, by measuring the period
and inclination, the remarkable masses of the WR~20a components.

In this paper, we present photometry for WR~20a, which clearly shows
the presence of eclipses. In Section 2 we describe the observations,
in Section 3 we present the light curve and the analysis and in
Section 4 we discuss the results.

\section{OGLE Observations}

WR~20a $\rm(=SMSP2=THA 35-II-036)$ is located in the constellation
Carina at $\alpha=10$:23:58.0, $\delta= -57$:45:49 (J2000.0).
Observations were carried out with the 1.3 m Warsaw telescope at Las
Campanas Observatory, Chile, which is operated by the Carnegie
Institute of Washington. The telescope is equipped with a mosaic CCD
camera with 8192$\times$8192 pixels. WR~20a is very bright, $I\sim
11.0$, so very short exposures of 10 seconds each were used to avoid
saturating the CCD.  The observations presented here started on May 1,
2004 and consist of 83 measurements obtained on 17 consecutive nights.

The absolute calibration of our photometry was obtained on one
photometric night with a couple of standards from Landolt (1992).  The
uncertainty in the zero points of the photometry is about 0.03 mag.

The data were reduced using the standard OGLE-III data pipeline, as
described by \citet{Udalski03}. The photometry was derived using
difference imaging analysis.  This method is the current
state-of-the-art for photometric accuracy in crowded fields
\citep{Alard00}. Our analysis resulted in a set of time-series
photometry in the $I$-band, presented in Table~1. The formal error in
the photometry is as low as $0.001-0.002\;$mag, but we have adopted a
conservative error of $0.01\;$mag for the differential light
curve. $V$-band images were also obtained on one night during
maximum. We measured the following magnitudes for WR~20a at maximum:
$V=13.45 \pm 0.025$, and $I=10.65 \pm 0.05$.

\section{Light Curve and Radial Velocity Curve Analysis}

The first goal of our observations was to confirm and refine the
orbital period of the WR~20a binary. As discussed extensively by
\citet{Rauw04}, besides the main period of $P=3.675\;$days, their
spectroscopic data also allowed for a much shorter period of
$1.293\;$days, which would result in reduced minimum masses of about
$25\; \msun$ for each star in the system.  A longer period of
$4.419\;$days was also possible, resulting in increased minimum masses
of above $80 \;\msun$ for each star. Using the analysis of variance
technique of \citet{Schwarzenberg89}, the most significant period
derived from our data is $P=1.843 \pm 0.005$ days, where the error is
a conservative 3$\sigma$ estimate. However, we know that the system
has to have two eclipses, which results in a true orbital period twice
as long. We find an ephemeris for the primary (deeper) eclipse of

\begin{equation}
\rm T_{prim}=2453124.569 + 3.686 \times E,
\end{equation}

\noindent thus confirming and refining the orbital period of WR~20a
and the remarkable masses of its components.

In addition, we can derive an inclination angle $i$ for the system
from our well-sampled light curve. To first order, the value of the
inclination angle is given by the depth of the eclipses ($\sim 0.4$
mag), and we find that the exact value of $\;i\;$ only weakly depends
on the details of the model fit, or even which eclipsing binary model
we use to fit. To demonstrate this, we first fit the light curve with
a simple model of two spherical stars with limb darkening (J. Devor
2004, private communication), as well as with the more complex EBOP
model \citep{Nelson72,Popper81} and finally, with the Wilson-Devinney
(WD) code \citep{Wilson71,Wilson79,vanHamme03} for modeling distorted
stars.  We find that indeed the inclination angle is insensitive to
the model, and the best fit value with WD is $i=74.5\pm 2.0\deg$. 

\begin{figure}[h]  
\plotone{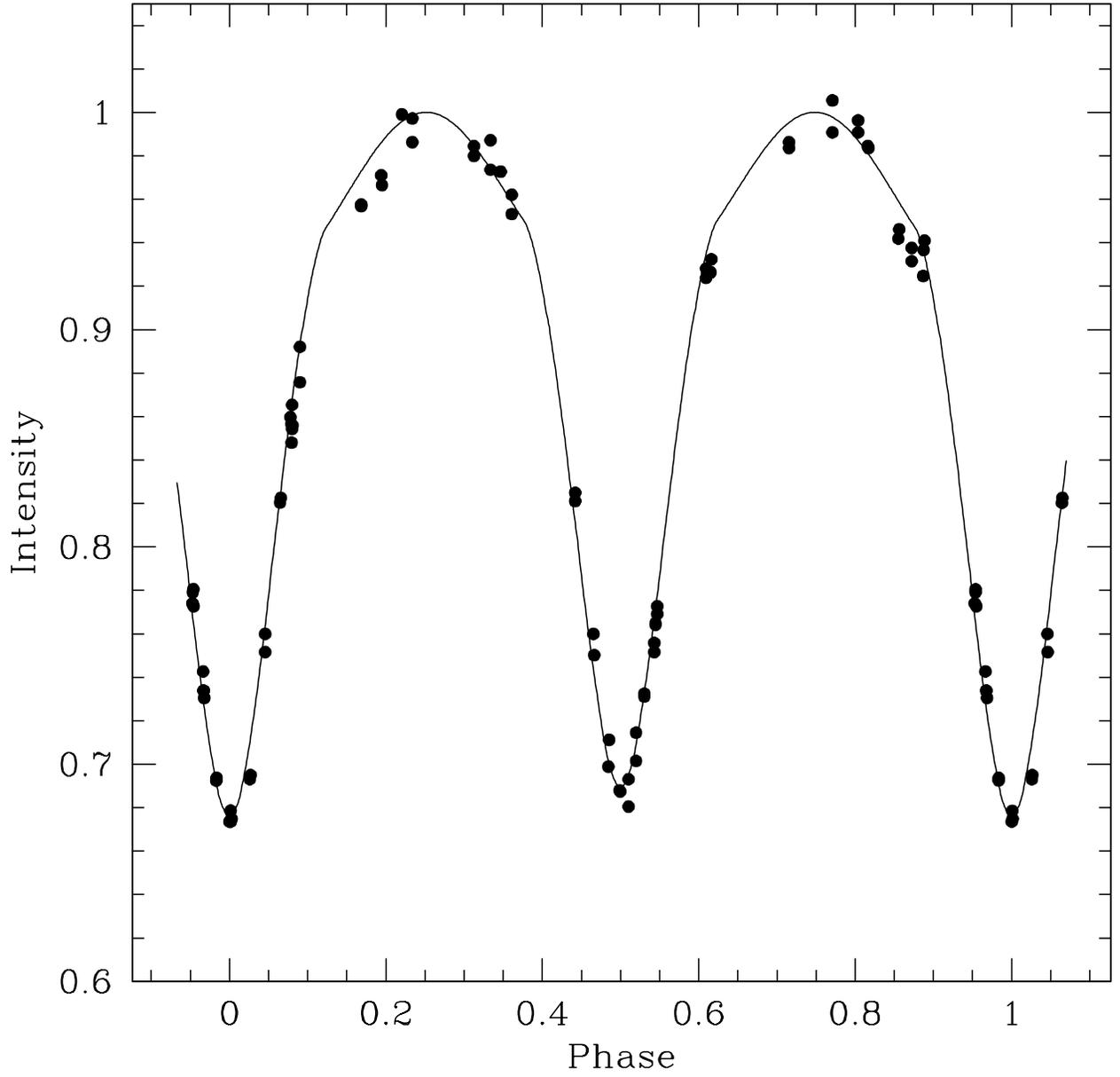}
\caption{Wilson-Devinney model fit of a near contact binary to the
$I$-band light curve of WR~20a. The period is $3.686$ days, the
eccentricity is $0$ and the inclination $i=74.5 \deg$.}
\label{lc}
\end{figure}

We ran WD in the overcontact mode (MODE 3), fixing T$_{\rm
eff1}=42\,000$K \citep[since WR 20a is intermediate between Mk42 and
WR 47c, see][]{Rauw04,Crowther98}, using linear limb darkening
coefficients from \citet{vanHamme03}, values for gravity darkening
exponents and albedos from theoretical values for radiative envelopes,
and a mass ratio determined from the radial velocity curve (see
below). We fit for the inclination $i$, T$_{\rm eff2}$, the luminosity
of the primary and the surface potential ($\Omega_{1}=\Omega_{2}$) and
defined convergence to be reached after five consecutive iterations
for which the corrections for all adjusted parameters were smaller
than their respective standard (statistical) errors. The results of
the fit are given in Table~2. In Figure~\ref{lc} we show the result of
the $I-$band light curve model fit for WR~20a. The uneven eclipse
depths suggest slightly different effective temperatures and, thus,
different spectral types for the components. Using differential
photometry we can measure a difference between the two stars as small
as 1-2\%, which is not possible with spectral decomposition. However,
multi-band photometry is necessary to resolve the effective
temperature -- radius degeneracy. Finally, we derive the following
best fit fractional radii for both stars: polar radius $\rm
r_{pole}=0.34$, radius to the inner Lagrangian point $\rm
r_{point}=0.40$, $\rm r_{side}=0.35$, and $\rm r_{back}=0.37$.

\begin{figure}[h]   
\plotone{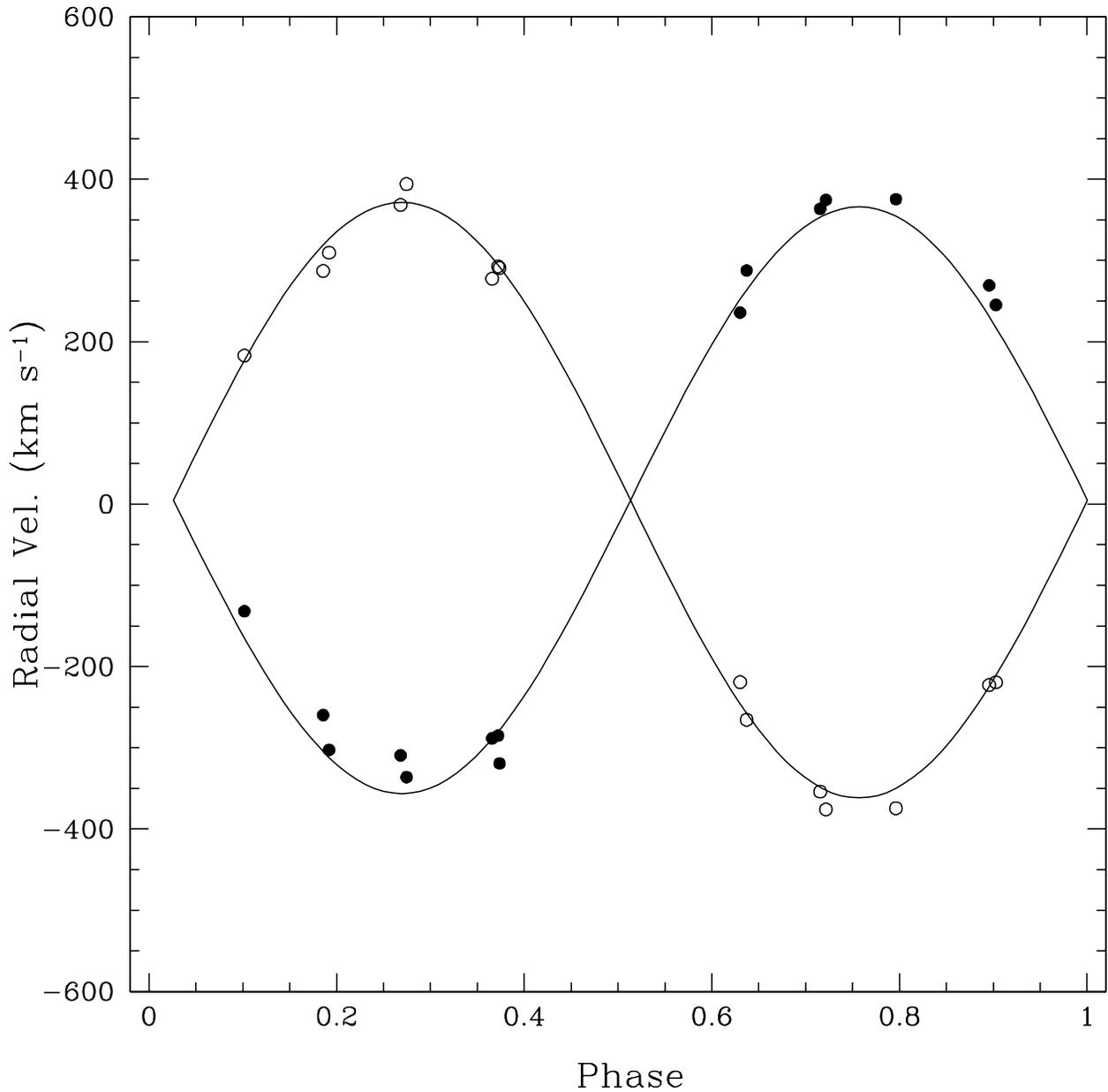}
\caption{Radial velocity curve fit to measurements of WR~20a made by
\citet{Rauw04}. The filled circles correspond to the primary (more
massive) star. The new semi-amplitude values are $\rm K_{1}=362\;\rm
km\; s^{-1}$ and $\rm K_{2}=366\;\rm km\; s^{-1}$ and the masses are
$\rm m_{1}\,sin^{3}\, \it i=\rm 74.3\, \msun$ and $\rm
m_{2}\,sin^{3}\, \it i= \rm 73.4\, \msun$, which, combined with
$i=74.5\deg$ yield mass values for the components of $83.0\; \msun$
and $82.0\; \msun$.}
\label{rv}
\end{figure}

Armed with the refined period and the exact value of the inclination
angle, we re-analyzed the radial velocity data of
\citet{Rauw04}. Fixing the eccentricity to $0$, as the light curve
confirms, and the period to $3.686$ days, we fit the radial velocity
data applying equal weights and rejecting points with radial velocity
measurements smaller than $80\;\rm km\;s^{-1}$. These measurements
were made by fitting Gaussians to lines that are not resolved and thus
include errors much larger than $10-15\;\rm km\; s^{-1}$ and are also
subject to non-Keplerian effects. Our fit is shown in Figure~\ref{rv},
and the best fit values with their statistical errors are presented in
Table~3. We note that due to the refined period, as \citet{Rauw04}
caution, their two earliest RV epochs are now ``flipped''. We derive
slightly larger velocity semi-amplitudes than \citet{Rauw04}, which in
turn yield larger masses. Specifically, the values are $\rm
K_{1}=362.2\;\rm km\; s^{-1}$ and $\rm K_{2}=366.4\;\rm km\; s^{-1}$,
the mass ratio $\rm q=m_{2}/m_{1}=0.99$, the systemic velocity
$\gamma=8\; \rm km\; s^{-1}$, the semi-major axis $\rm a \;sin\, \it
i=\rm 53\; \rsun$, and finally the masses $\rm m_{1}\;sin^{3}\, \it
i=\rm 74.3\; \msun$ and $\rm m_{2}\;sin^{3}\, \it i= \rm 73.4\;
\msun$. Knowing the inclination, we calculate the masses of the stars
to be $83.0 \pm 5.0\; \msun$ and $82.0 \pm 5.0\; \msun$. These erros
include the error in the minimum mass and the error in the
inclination.

\section{Discussion}

As discussed at length by \citet{Rauw04}, the extreme mass values of
WR~20a make it a ``cornerstone system for future investigations of
massive star evolution''. We confirm and refine these extreme values,
by measuring the masses of the components to be $83.0 \pm 5.0\; \msun$
and $82.0 \pm 5.0\; \msun$. If the WN6ha spectral type is confirmed,
WR~20a will be one of few known WR binaries with a mass ratio near
unity. Thus, it becomes very important now to obtain high resolution
spectra and multiband photometry of WR 20a, corrected for eclipses.

Such future observations will allow an accurate reddening measurement
and furthermore, a direct, independent and accurate distance
determination to the system. This will resolve the discrepancy in the
distance to Westerlund 2, which ranges from $2.5-8$ kpc \citep[see
discussion in][]{Churchwell04}. Several multiband measurements of
WR~20a already exist. As measured by us (Section 2), WR~20a has
$V=13.45$ and $I=10.65$ at maximum. Our $V$ is different from
$V=13.58$ measured by \citet{Moffat91}, but their measurement could
have been affected by the eclipses. The Two Micron All Sky Survey
\citep[{\it 2MASS\rm},][]{Skrutskie97} has measured $J=8.86, H=8.08$
and $K=7.59$ for WR~20a. Recent infrared observations of the HII
region RCW 49 with the $Spitzer\; Space\; Telescope$ \citep{Whitney04}
contain measurements of the binary. \citet{Belloni94} have obtained
$ROSAT$ observations of RCW 49, tentatively identifying WR 20a with
one of their point sources. However, further observations are needed
to determine the extinction to WR~20a.

Analogs of WR~20a, possibly with even higher mass components, are
bound to exist in nearby galaxies, such as M31 and M33. There, they
would be first found by variability searches such as DIRECT
\citep{Stanek98,Bonanos03} as bright eclipsing binaries (about 18th
magnitude for $\rm M_{V}=-6.5$), which can then be followed by
spectroscopy to obtain radial velocities. It is our experience
(A.Z. Bonanos et al., in preparation) that accurate radial velocities
can be obtained for such systems with the existing 6.5-10 meter
telescopes. The advantage of studying eclipsing binaries in these
galaxies would be their well known distances and low reddening, as
systems with high reddening would be too faint to follow-up.

We conclude that WR~20a truly is a ``cornerstone system'' that
deserves detailed additional studies. Not only will studies of the
most massive binaries constrain models, but also will help understand
other problems, such as the nature of mass loss and winds of early
type stars, as well as the connection of these stars to supernovae and
gamma ray bursts \citep{Stanek03,Maund04}. WR~20a now overtakes the
previous record holders WR 22 \citep{Rauw96,Schweickhardt99} and
R136-38 \citep{Massey02} in the most massive star competition, by
becoming ``the most massive star ever weighed.''

\acknowledgments{We thank G. Torres for a careful reading of and
comments on the manuscript and D. Sasselov and J. Devor for useful
discussions. We also thank the referee, Eric Gosset, for his prompt
and careful reading of the manuscript and his useful comments. AZB and
KZS were partially supported by HST Grant HST-GO-09810.06-A. Support
for Proposal number HST-GO-09810.06-A was provided by NASA through a
grant from the Space Telescope Science Institute, which is operated by
the Association of Universities for Research in Astronomy,
Incorporated, under NASA contract NAS5-26555. Partial support to the
OGLE project was provided with the following grants: Polish KBN grant
2P03D02124 to A. Udalski and NSF grant AST-0204908 and NASA grant
NAG5-12212 to B.~Paczy{\'n}ski.  }

%%%%%%%%%%%%%%%% BIBLIOGRAPHY  %%%%%%%%%%%%%%%%%%%%%%%%
\clearpage
\bibliographystyle{apj}
\bibliography{ref}
\clearpage
\input{table1.tex}
\input{table2.tex}
\input{table3.tex}
\end{document}

%% file: table1.tex
\begin{deluxetable}{clcc}
\tabletypesize{\footnotesize}
\tablewidth{0pc}
\tablecaption{\sc $I-$band Photometry of WR 20$a$}
\tablehead{
\colhead{UT} &\colhead{HJD} & \colhead{$I$} & \colhead{$\sigma_{\rm I}$}} 
\startdata
May 1, 2004 &  2453126.57440 & 10.932 &   0.01 \\
......  &  2453126.57581 & 10.931 &   0.01 \\
......  &  2453126.58207 & 10.920 &   0.01 \\
......  &  2453126.58350 & 10.925 &   0.01 \\
May 2, 2004 &  2453127.57488 & 10.657 &   0.01 \\
......  &  2453127.57883 & 10.658 &   0.01 \\
May 3, 2004 &  2453128.54163 & 10.804 &   0.01 \\
......  &  2453128.54306 & 10.808 &   0.01 \\
......  &  2453128.54659 & 10.811 &   0.01 \\
......  &  2453128.54800 & 10.797 &   0.01 \\
May 4, 2004 &  2453129.58401 & 10.682 &   0.01 \\
....... &  2453129.58545 & 10.692 &   0.01 \\
May 5, 2004 &  2453130.52109 & 10.723 &   0.01 \\
......  &  2453130.52253 & 10.716 &   0.01 \\
May 6, 2004 &  2453131.52484 & 10.711 &   0.01 \\
......  &  2453131.52627 & 10.706 &   0.01 \\
May 7, 2004 &  2453132.55993 & 10.687 &   0.01 \\
......  &  2453132.56136 & 10.688 &   0.01 \\
......  &  2453132.65164 & 10.672 &   0.01 \\
......  &  2453132.65307 & 10.677 &   0.01 \\
......  &  2453132.75066 & 10.641 &   0.01 \\
May 8, 2004 &  2453133.56632 & 10.849 &   0.01 \\
......  &  2453133.56775 & 10.854 &   0.01 \\
......  &  2453133.65146 & 10.938 &   0.01 \\
......  &  2453133.65288 & 10.952 &   0.01 \\
......  &  2453133.72505 & 11.029 &   0.01 \\
......  &  2453133.72649 & 11.010 &   0.01 \\
May 9, 2004 &  2453134.57341 & 10.655 &   0.01 \\
......  &  2453134.57483 & 10.658 &   0.01 \\
May 10, 2004 & 2453135.45134 & 10.918 &   0.01 \\
......  &  2453135.45275 & 10.911 &   0.01 \\
......  &  2453135.45523 & 10.909 &   0.01 \\
......  &  2453135.45666 & 10.920 &   0.01 \\
......  &  2453135.50139 & 10.963 &   0.01 \\
......  &  2453135.50284 & 10.976 &   0.01 \\
......  &  2453135.50518 & 10.976 &   0.01 \\
......  &  2453135.50660 & 10.981 &   0.01 \\
......  &  2453135.56331 & 11.039 &   0.01 \\
......  &  2453135.56473 & 11.037 &   0.01 \\
......  &  2453135.62564 & 11.069 &   0.01 \\
......  &  2453135.62707 & 11.069 &   0.01 \\
......  &  2453135.62904 & 11.061 &   0.01 \\
......  &  2453135.63045 & 11.067 &   0.01 \\
......  &  2453135.72198 & 11.038 &   0.01 \\
......  &  2453135.72339 & 11.035 &   0.01 \\
May 11, 2004 &  2453136.48681 & 10.655 &   0.01 \\
......  &  2453136.48824 & 10.643 &   0.01 \\
May 12, 2004 &  2453137.46474 & 11.046 &   0.01 \\
......  &  2453137.46618 & 11.047 &   0.01 \\
......  &  2453137.50263 & 11.038 &   0.01 \\
......  &  2453137.50407 & 11.058 &   0.01 \\
......  &  2453137.53984 & 11.025 &   0.01 \\
......  &  2453137.54127 & 11.010 &   0.01 \\
......  &  2453137.58009 & 10.980 &   0.01 \\
......  &  2453137.58149 & 10.978 &   0.01 \\
......  &  2453137.62247 & 10.950 &   0.01 \\
......  &  2453137.62389 & 10.944 &   0.01 \\
May 13, 2004 &  2453138.46632 & 10.634 &   0.01 \\
......  &  2453138.46773 & 10.650 &   0.01 \\
......  &  2453138.58603 & 10.644 &   0.01 \\
......  &  2453138.58744 & 10.650 &   0.01 \\
May 14, 2004 &  2453139.48039 & 10.938 &   0.01 \\
......  &  2453139.48180 & 10.950 &   0.01 \\
......  &  2453139.54939 & 10.855 &   0.01 \\
......  &  2453139.55083 & 10.852 &   0.01 \\
......  &  2453139.60100 & 10.819 &   0.01 \\
......  &  2453139.60241 & 10.809 &   0.01 \\
......  &  2453139.63839 & 10.764 &   0.01 \\
......  &  2453139.63982 & 10.784 &   0.01 \\
May 15, 2004 &  2453140.46213 &  10.662 &   0.01\\
......  &  2453140.46355 &  10.657 &   0.01\\
......  &  2453140.53695 &  10.654 &   0.01\\
......  &  2453140.53839 &  10.669 &   0.01\\
......  &  2453140.58621 &  10.670 &   0.01\\
......  &  2453140.58762 &  10.670 &   0.01\\
May 16, 2004 &  2453141.55358 &  10.726 &   0.01\\
......  &  2453141.55499  & 10.721 &   0.01\\
May 17, 2004 &  2453142.46556 &  10.705 &   0.01\\
......  &  2453142.46700 &  10.700 &   0.01\\
......  &  2453142.52384 &  10.710 &   0.01\\
......  &  2453142.52525 &  10.717 &   0.01\\
......  &  2453142.57798 &  10.725 &   0.01\\
......  &  2453142.57940 &  10.725 &   0.01\\
\enddata
\end{deluxetable} 

%% file: table2.tex
\begin{deluxetable}{ll}
\tabletypesize{\footnotesize}
\tablewidth{0pc}
\tablecaption{\sc Light Curve Parameters for WR 20$\rm a$}
\tablehead{
\colhead{Parameter} & \colhead{Value}} 
\startdata
Period, P & 3.686 $\pm$ 0.01 days \\
Primary eclipse, T$_{\rm prim}$ & 2453124.569 \\
Inclination, $i$	& 74.5 $\pm$ 2.0 $\deg$\\
Eccentricity, $e$ & 0 (fixed) \\
Effective Temperature, T$_{\rm eff1}$& 42000 K (fixed) \\
Effective Temperature, T$_{\rm eff2}$& 40300 $\pm$ 1000 K\\
Surface Potential ($\Omega_{1}=\Omega_{2})$ & 3.92 $\pm$ 0.03 \\
Radius, $\rm r_{pole}$   & 18.7 $\pm$ 0.3 $\rsun$ \\     
Radius, $\rm r_{point}$  & 22.0 $\pm$ 0.3 $\rsun$ \\
Radius, $\rm r_{side}$   & 19.3 $\pm$ 0.3 $\rsun$ \\
Radius, $\rm r_{back}$   & 20.4 $\pm$ 0.3 $\rsun$ \\
\enddata
\end{deluxetable} 

%% file: table3.tex
\begin{deluxetable}{ll}
\tabletypesize{\footnotesize}
\tablewidth{0pc}
\tablecaption{\sc Radial Velocity Curve Parameters for WR 20$\rm a$}
\tablehead{
\colhead{Parameter} & \colhead{Value}} 
\startdata
Systemic velocity, $\gamma$ & 8.0 $\pm$ 5.0 $\rm km\;s^{-1}$ \\
Semi-amplitude, K$_{1}$ & 362.2 $\pm$ 8.0 $\rm km\;s^{-1}$ \\
Semi-amplitude, K$_{2}$ & 366.4 $\pm$ 8.0 $\rm km\;s^{-1}$ \\
Semi-major axis, a sin$\;i$ & $53.0\;\pm 1.0 \; \rsun$\\
Mass Ratio, q & 0.99 $\pm$ 0.03 \\
Mass, m$_{1}\; \rm sin^{3} \it i$ & $74.3\; \pm 4.0\; \msun$ \\
Mass, m$_{2}\; \rm sin^{3} \it i$ & $73.4\; \pm 4.0\; \msun$ \\
\enddata
\end{deluxetable} 